\documentclass[a4paper,12pt,authoryear]{article}

\usepackage{amstext}
\usepackage{amsthm}
\usepackage{hyperref}
\usepackage{graphicx,graphics,multirow,epstopdf}
\usepackage{amsmath,amsfonts,amssymb}
\usepackage{fancyhdr}
\usepackage{booktabs}
\usepackage{subcaption}
\usepackage{xcolor}

\usepackage{natbib}
\newcommand{\blind}{0}

\begin{document}

\date{\today}

\if0\blind
{

\title{Multivariate Constrained Robust M-Regression for Shaping
Forward Curves in Electricity Markets}
\author{Peter Leoni \\
Department of Mathematics, KU Leuven, Belgium\\
Pieter Segaert\\
Department of Mathematics, KU Leuven, Belgium\\
Sven Serneels\\
BASF Corporation, Tarrytown, NY 10591, USA\\
Tim Verdonck \footnote{Corresponding author: Tim Verdonck,  Celestijnenlaan 200B, 3001 Leuven, Belgium,  tel. +32 16 32 27 89, fax +32 16 3 27998, mail: tim.verdonck@kuleuven.be}
\thanks{This research has benefited from the financial support of the Flemish Science Foundation (FWO) and project C16/15/068 of Internal Funds KU Leuven.
}
\\
Department of Mathematics, KU Leuven, Belgium\\
}
\maketitle
} \fi

\if1\blind
{
  \bigskip
  \bigskip
  \bigskip
  \begin{center}
    {\bf Multivariate Constrained Robust M-Regression for Shaping
Forward Curves in Electricity Markets}
\end{center}
  \medskip
} \fi
\bigskip

\begin{abstract}
In this paper, a multivariate constrained robust M-regression (MCRM) method is developed to estimate shaping coefficients for electricity forward prices. An important benefit of the new method is that model arbitrage can be ruled out at an elementary level, as all shaping coefficients are treated simultaneously.  
Moreover, the new method is robust to outliers, such that the provided results are stable and not sensitive to isolated sparks or dips in the market. 
An efficient algorithm is presented to estimate all shaping coefficients at a low computational cost. To illustrate its good performance, the method is applied to German electricity prices.
\end{abstract}
\vfill

\newpage

\section{Introduction}

It is a well known fact that energy forward and future prices, such as 
natural gas or electricity prices are highly seasonal \citep{Geman05,Huisman09}.
This is due to various factors such as weather dependent supply and demand 
and the problem of energy storage \citep{Weron07,BoogertDupont06}. In 
contrast to a forward in classical financial markets, a forward in energy 
refers to a contract that provides the delivery of the underlying commodity 
over a fixed delivery period \citep{Schofield07}. This can be anything from 
a block of 15 minutes to a full year, depending on the contract.

In Europe, the forward market has developed into a cascading series of
prices, whereby the close-to-delivery part of the curve is more densely
populated by forward contracts with higher or finer granularity such as
days, weekends, weeks or months and the further-from-delivery forwards are
only being traded in the form of quarterly, seasonally or yearly contracts.

The problem of transforming prices of traded contracts with a low
granular nature into high granularity contracts has been treated by various authors. The work of \cite{fleten2003constructing} may be seen as the seminal work on this topic. They propose to obtain an average seasonal shape curve using regression techniques. This seasonal curve can then be used to estimate prices for finer granularity contracts from higher granularity contracts. For the regression model, one typically considers dummy variables for the weekly and yearly cycles, as well as information on the cooling and heating degree days (see e.g. \cite{hildmann2012combined, kiesel2018construction}).  However, this regression approach typically induces arbitrage in the model. As a solution for this problem, \cite{fleten2003constructing} propose to adjust the curves obtained in the previous step by smoothing them and imposing a non-arbitrage condition. 

The work of \cite{fleten2003constructing} motivated various authors to consider more advanced models. \cite{koekebakker2004modelling} and \cite{benth2007extracting} propose a combination of seasonal paths and fourth order polynomial splines using maximum smoothness interpolation introduced by \cite{adams1994fitting}. These approaches are, however, quite elaborate and more involved to apply in practice. Moreover, \cite{Borak08} reported these methods to be sensitive to model risk. Instead they propose a dynamic semi-parametric factor model. On the other hand \cite{caldana2016electricity} noted that the algorithm of \cite{Borak08} suffers from underfitting market prices and may fail to account for short-term periodical patterns. A comparison between various approaches including works by \cite{fleten2003constructing}, \cite{benth2007extracting} and \cite{paraschiv2016estimation} may be found in \cite{kiesel2018construction}. 

The above methods are all non-robust in the sense that they try to fit an optimal model for all observations. Therefore, they are highly susceptible to the possible presence of atypical observations or outliers in the data. As the non-robust fit of the model is attracted by the outliers, these observations may no longer appear as outliers after the fit. This effect is known as masking. In the worst possible scenario the effect of outliers on a non-robust fit can be so explicit that regular observations appear to be outlying. This effect is called swamping. Both concepts were illustrated by \cite{davies1993identification}. It is important to note that any detected outlier need not necessarily be an error in the data. Their presence may reveal that the data is more heterogeneous than assumed. Outliers may also come in clusters, indicating there are subgroups in the population that behave differently. A robust analysis can thus provide better insights in the structure of the data and reveal structures in the data that would remain hidden in a classical analysis. Extensive literature exists on detecting outliers and developing methods that are robust to them. An overview may be found in \cite{Rousseeuw86-03} and \cite{Maronna}. 

The problem of robustness for forward curves was also noted by \cite{hildmann2011robust} and \cite{hildmann2012combined} who used the LAD to obtain a robust estimate of seasonality shape. However, they need a separate step to solve the arbitrage question and their methodology focuses on the hourly price forward curve. Also \cite{caldana2016electricity} considered some notions of robustness in their proposed model. As a downside, the outlier detection rule they propose uses estimated standard deviations which are highly susceptible to outliers on their turn as well.

In this paper, a statistically sound and robust method is proposed that allows to treat both the higher and fine granular data, as well as the tradeable forward directly, so as to establish a consistent market pattern. The proposed methodology is applicable to any level of granularity. The arbitrage question is incorporated directly in the estimator, therefore obliterating the need for additional separate steps. Moreover, the methodology is inspired by how the forward market trades, rather than trying to translate time-series on spot like data (high granular) into long term patterns, where true seasonality can be hidden behind noise trends and volatility. Motivated by these considerations, a multivariate constrained robust regression (MCRM) technique is developed that calculates fast on large data sets and is relatively easy to explain to practitioners. Note that the method can be seen as a building step in the overall process and that potentially, filtering techniques can further improve its quality. 

In Section \ref{Sec:markets} the electricity market and its typical properties related to forward prices are briefly described. 
Section \ref{Sec:traditional} discusses the usage of shaping coefficients in electricity markets. A new method for shaping forward curves in electricity markets is proposed in Section \ref{Sec:proposal} and the development of a multivariate constrained robust M-regression for these purposes, is motivated. In Section \ref{Sec:algo}, the estimator is described in more detail and an efficient algorithm is given in pseudo-code.  The proposed methodology is applied on real data from the German power market in Section \ref{Sec:example} to illustrate its performance. In this section, the proposed method is also compared to the method of \cite{fleten2003constructing}, assessing both the estimation and prediction performance of both methods. Finally, some conclusions and possible outlook are given in Section \ref{Sec:conclusion}.

\section{Forward prices in electricity markets\label{Sec:markets}}

This paper will focus on the European electricity
market \citep{Huisman09,Bunn04}, but the proposed methodology is of course also applicable and
relevant in other electricity markets. 

Throughout the article, there are three common ways of referring to forward contract prices. The price can be
denoted by $F\left( t,T_{1},T_{2}\right) $ where $t$ stands for the
reference or observation date and $\left[ T_{1},T_{2}\right] $ denotes the
delivery period, e.g. $T_{1}=1/1/2013$ and $T_{2}=31/12/2013.$ Alternatively,
$F\left( t,\delta \right) $ denotes the same contract, where $%
\delta $ denotes the delivery period as a whole, for example $Cal-2013$. In some
cases, it is easier to use relative delivery periods, for which the notation $F\left( t,\rho \right) $ is being reserved, with $\rho $ denoting a
relative delivery period such as $Y+1$ (one year forward) with respect to
the observation date $t$. From the context it will always be clear which
notation is being used.

An important relationship that one has to understand in electricity markets, is
how the contracts relate to each other. Purchasing a baseload contract for delivery
in $Q_3-2012$ (the third quarter of the year 2012) at a forward price of $41.65$ EUR/MWh in a volume of $10$MW
implies that the holder of this contract will be delivered $10$MW of power
for every single hour, starting 1/7/2012 at midnight, until midnight on
30/9/2012. It is clear that holding 3 separate contracts for power delivery
in July-12, August-12 and September-12, each for the volume of $10$MW is an
equivalent position. This implies that the prices of these 4 contracts are
related to each other by a non-arbitrage constraint. Since all prices are
quoted as average prices over their delivery period (e.g. EUR/MWh), one can
loosely say that the average over the monthly contracts has to equal the
price of the quarterly contract. The exact relationship depends on the
settlement or payment dates of the forward, the appropriate discount factors
and the exact number of days within each delivery month.

The amount of tradeable contracts highly depends on the market and several types contracts might be available. As en example, different types of forward contracts are listed for German Baseload Power that were being traded on the third of May, 2012. For example, several short contracts such as D+1 (May 4rd), the following weekend (WE+1) of the following week (W+1) were traded. But also longer contracts for the next month (M+1), quarter (Q+1) or year (Y+1) were traded. In general one can expect to have at least some of the type of contracts of Table \ref%
{TableSnapShot} to be available. In illiquid markets this table will reduce
to a few entries such as the day-ahead price and some calendar contracts. This paper will focus on the German power market, which
has a good liquidity because of the vast number of participants such as
banks, hedge funds, utilities, but also large industrials.

\begin{table}[!tb]
\centering
\begin{tabular}{ll|ll|ll}
\toprule
Period & Price & Period & Price & Period & Price \\ \midrule
D+1 & 44.75 & M+1 & 41.45 & Q+1 & 43.20 \\ 
D+2 & 39.00 & M+2 & 42.40 & Q+2 & 52.85 \\ 
 &  & M+3 & 41.00 & Q+3 & 54.40 \\ 
WE+1 & 36.60 & M+4 & 46.50 & Q+4 & 45.10 \\ 
WE+2 & 33.50 &  &  & Q+5 & 45.80 \\ 
 &  &  &  & Y+1 & 50.20 \\ 
W+1 & 43.00 &  &  & Y+2 & 50.20 \\ 
W+2 & 38.50 &  &  & Y+3 & 50.50 \\ \bottomrule
\end{tabular}%
\caption{\label{TableSnapShot} The list of traded forward contracts for German Baseload Power on
3rd May 2012. All prices are expressed in Eur/MWh.}
\end{table}

\begin{figure}[!h]
\begin{center}
\includegraphics[width=9cm]{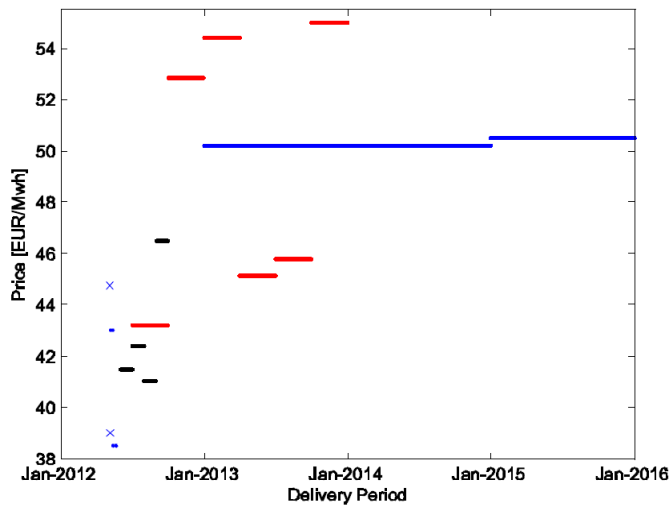}
\caption{\label{Fig:GBP} The traded forward contracts for German Base Power on August 24, 2011. The $x$ axis represents the delivery period and the $y$ axis the average price over the delivery period.}
\end{center}
\end{figure}

Another factor of electricity prices are seasonal patterns. This is for the quarterly prices of calendar year 2013 in the German market in Figure~\ref{Fig:GBP} (observation date August, 24 2011). When moving further
down the forward curve into 2014 and 2015, similar
seasonal market information is absent. There, the market is not liquid enough and
although the fundamentals of the electricity market predict that seasonal
patterns will be there as soon as the calendar contract cascades into
quarters, they cannot be observed at that time.

Since valuation of long term contracts and asset valuation often largely
depend on the specific price levels of the quarters, rather than on the
average calendar level, the common approach in the market is so-called
\textit{shaping} of the curve. This means one tries to use relevant historical
patterns in order to ``guesstimate'' the shape of the quarter around the
calendar. This can be done based on economical principles (fundamental
approach), such as predictions on the available production capacity of power,
taking into account new plants that will come online or old plants that get
shut down. However, a reliable estimate would at least take into account
market information of the (recent) past. In its most rudimentary form, one
could try and copy the shape one has for 2012/2013 onto the consecutive
years.

For some valuation problems, one needs to drill down to hourly price
patterns in order to assess the value of a contract. A typical power market is subject to various degrees of seasonality. On the low
granular side, one can distinguish the shape of the seasons, quarters or
months versus the average price levels in a year. On a finer granularity,
one can observe weekly patterns where weekends typically have a lower price
than weekdays. But one can drill even further down and observe price
patterns within a given day where it is normal to expect more demand during
daytime than during nighttime. In fact these patterns are really
specific to the supply and demand profile, reflecting the rhythm of the day and vary
themselves across the year.

\section{Estimation of shaping coefficients for electricity forward prices\label{Sec:traditional}}

Most academic research on modeling electricity prices either brings in a
more economic or fundamental approach \citep{Bunn04,SkantzeIlic12} or chooses
to focus on the stochastic nature of the price curve \citep{ClewlowStrickland00, Fiorenzani06,Pilipovic07}. The fundamental models aim
to explain price patterns or changes within price regimes. This approach
brings a valuable qualitative understanding of how electricity prices are formed
and what the price drivers are. A recent issue that has arisen, is 
how the phase-out of the nuclear and the increase of renewable capacity will influence the price levels in
Europe. 

The stochastic models are key to perform valuation of derivatives contracts
in the energy markets such as tolling agreements, gas storages or long-term
gas swing contracts \citep{EydelandWolyniec03,James12}. However, an important aspect in derivatives pricing that is often overlooked, is the effect of the assumed highly granular seasonality. In fact,
for far ahead price contracts, the seasonal pattern is not visible in the
market and one has to rely on assumptions or a model \citep{Koekebakker07} to
derive a curve with a high enough granularity all across the curve. This
curve in effect forms the basis for the valuation process and the stochastic
model is built on top of this curve to describe the volatility of the price
curve. Since these contracts with high granularity on the back-end are not
traded, there is little information on the market consensus and classical
mathematical finance fails to give the correct answers, as there is no way to hedge or trade the risk.

It can easily be verified that the effect of the assumptions on
the seasonal shape of the curve are equally (or even more) important to the
valuation of any complex structure. Moreover, different assumptions on
the model can easily lead to bigger discrepancies than the uncertainty on
the volatility would bring.

In order to construct forward curves we define Year-to-Quarter (YtQ) coefficients, Quarter-to-Month (QtM)
coefficients, Month-to-Day (MtD) coefficients and Day-to-Hour (DtH) coefficients. Each of these coefficients transforms a lower granular forward price into a curve of higher granularity and have a fundamental and natural meaning attached to them. For example, the MtD coefficients give a reflection of the seasonality on a weekly basis, or more in particular the weekend versus weekday patterns that are typically present in electricity prices.  By decomposing the process in separate steps, one can keep more control over the assumptions and the effects on the final result. 

The usage of these coefficients is in line with the work of \cite{fleten2003constructing}. Indeed in the first step of this method a regression model with dummy variables is used to break up low granular price curves into high granular price curves. A detailed discussion may be found in \cite{hildmann2012combined} and \cite{kiesel2018construction}. In contrast to these works, our methodology will include the non-arbitrage condition in the estimation framework and will be robust. 

Based on the coefficients defined above, one may then assume a scaling
ratio between prices of different granularities
\begin{equation}
F\left( t,\delta ^{\prime }\right) =F\left( t,\delta \right) \cdot \beta
\left( \delta ,\delta ^{\prime }\right) ,
\end{equation}%
where $\delta $ is a contract with lower granularity compared to $\delta
^{\prime }.$

The first step of the shaping process,
namely the conversion of yearly forward prices into quarterly prices, illustrates how scaling between different levels of granularity works. For
example, the model prices for each quarter in $2014$ can be derived from the
four YtQ shaping coefficients:%
\begin{equation}
\left\{ 
\begin{array}{lcl}
F\left( t,Q_1-14\right)  & = & F\left( t,Cal-14\right) \cdot \beta
_{1}^{YtQ} \\ 
F\left( t,Q_2-14\right)  & = & F\left( t,Cal-14\right) \cdot \beta
_{2}^{YtQ} \\ 
F\left( t,Q_3-14\right)  & = & F\left( t,Cal-14\right) \cdot \beta
_{3}^{YtQ} \\ 
F\left( t,Q_4-14\right)  & = & F\left( t,Cal-14\right) \cdot \beta
_{4}^{YtQ}%
\end{array}%
\right.  , \label{ShapingModel}
\end{equation}
with $F\left( t,Cal-14\right)$ the price for a 2014 calender year contract observed at time $t$.

\begin{figure}[!h]
\begin{center}
\includegraphics[width=9cm]{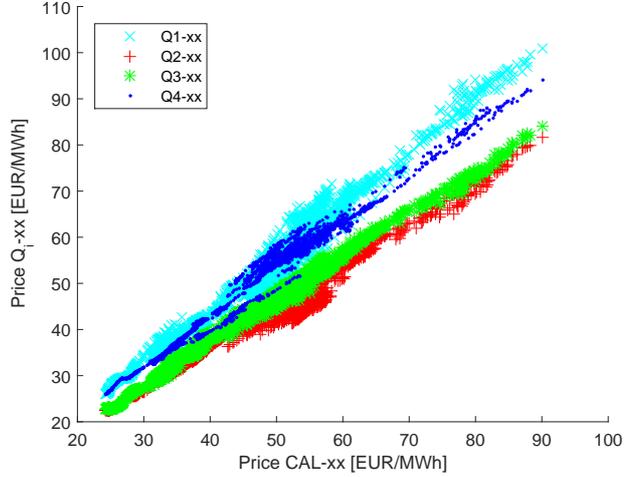}
\caption{\label{Fig:data} The quotation prices for each of the four quarters versus the calender year price. The data contains $Q_i-j(t_k)$ for all quotation dates $k=1,\ldots, N_{i,j}$, each quarter $i=1,\ldots, 4$ and all forward years $2004,\ldots, 2015$.}
\end{center}
\end{figure}

As an illustration of model $\left( \ref{ShapingModel}\right)$, real data from the German power market\footnote{obtained from the EEX exchange: www.eex.com} is used, which is the main power market in continental Europe. In Figure \ref{Fig:data} the quotation prices for each of the four quarters versus the calendar year price for all forward years from 2004 until 2015 are plotted. 

Despite large fluctuations in the prices over the different years, an underlying linear relationship for the whole data set is strikingly present. This supports the usage of shaping coefficients as defined above. Note that intercepts in this linear relation need not necessarily be zero.  Based on trading experience, the linear fit typically holds. Even if the delivery periods are split out over several years, a linear relationship is still clearly visible for most data. 

Note that the entire data history is used for illustration purposes only. From a fundamental point of view, taking this much history into account is not very sensible because the shapes in the market are undergoing changes with the introduction of solar and wind energy and the phase out of
the nuclear plants. An important advantage of a shaping method such as model $\left( \ref%
{ShapingModel}\right)$ is that analysts and traders can easily adjust
historical fits based on this new information since all parameters have
a natural interpretation. 

Denote the quarterly prices in the German power market data as follows: $F\left( t_{k},Q_{i}-j(t_{k})\right)$,  where $%
t_{k}$ stands for a quotation date and $k=1,\ldots ,N_{ij},$ $i=1,\ldots ,4$
for the specific quarter and $j=2004,\ldots ,2015$ for the historical years.
Similarly, the calendar prices are denoted as $F\left( t_{k},j\left(
t_{k}\right) \right) .$

By analyzing $\left( F\left( t_{k},Q_{i}-j(t_{k})\right) /F\left(
t_{k},j\left( t_{k}\right) \right) \right) _{j,k},$ one can estimate the
shaping coefficients $\beta _{1}^{YtQ},\ldots ,\beta _{4}^{YtQ}$ by means of
a simple average. If one uses all data and estimates the coefficients $%
\beta $ as the mean value over the data set, one obtains the following results%
\begin{equation}
\begin{array}{ccc}
\beta _{1}^{YtQ}=1.0926 & ; & \beta _{3}^{YtQ}=0.9398 \\ 
\beta _{2}^{YtQ}=0.8994 & ; & \beta _{4}^{YtQ}=1.0689%
\end{array}%
\label{coef:trad}
\end{equation}%
The weighted average of these coefficients is $1.0002,$ which introduces
a very slight arbitrage into the model, as in reality, this should be equal to one. 
Note that it is also possible to obtain an average shape that is smaller than one, hence the arbitrage can go both ways. For example, when
performing the same fit only on forward prices referring to Cal-12 and
Cal-13 only, one finds an average shape equal to $0.9994,$ showing that the
arbitrage can go both ways. 

The next section  presents a statistically sound method, which is applicable for every step of the shaping procedure and can still be grasped from an intuitive point of view. Moreover it naturally encompasses the non-arbitrage conditions, solving the issues found above in the simple averaging method.

\section{The concept of constrained robust regression for shaping forward curves\label{Sec:proposal}}


Resuming the example of the previous section, let $X$ denote the yearly forward price of a certain contract in the data. The goal is to obtain shaping coefficients $A_{k}$ with $k=1,\ldots,4$ transforming the yearly forward price into quarterly forward prices. Let $Y_{Q_k}$ denote the quarterly prices corresponding to this contract in the data. One then obtains the following joint set of equations:
\begin{align*}
Y_{Q_1} &= A_1 X + B_1 + \varepsilon_1 \\ 
Y_{Q_2} &= A_2 X + B_2 + \varepsilon_2 \\
Y_{Q_3} &= A_3 X + B_3 + \varepsilon_3 \\
Y_{Q_4} &= A_4 X + B_4 + \varepsilon_4
\end{align*}
with $B_k$ denoting the intercepts and $\varepsilon_k$ the error terms. To ensure the model is arbitrage free, an additional set of equations describing the relation between the coefficients has to be imposed. Define the columnvector $\gamma$ as follows: 
\begin{equation*}
\gamma^t =(A_{1},B_{1},A_{2},B_{2},A_{3},B_{3},A_{4},B_{4}).
\end{equation*}
The non-arbitrage condition is then imposed by specifying
\begin{equation*}
\left[ 
\begin{array}{cccccccc}
1 & 0 & 1 & 0 & 1 & 0 & 1 & 0 \\ 
0 & 1 & 0 & 1 & 0 & 1 & 0 & 1%
\end{array}
\right] \cdot \gamma = 
\left[ 
\begin{array}{cccccccc}
1 \\ 
0 
\end{array}
\right]. 
\end{equation*}
The general model is defined as:
\begin{equation}
Y_{k}=A_{k}\cdot X+B_{k} + \varepsilon_k,\, k \in [1,K],
\end{equation}%
with $K$ the number of subdivisions of the forward price X, $A_k$ the shaping coefficients and $B_k$ the intercepts. Just as before, the vector gamma is defined as $\gamma^t =(A_{1},B_{1},\ldots ,A_{K},B_{K})$. The non-arbitrage conditions are imposed trough the following set of equations: 
\begin{equation}
A_{eq}\cdot \gamma =b_{eq}.  \label{Ref:Equality}
\end{equation}
with 
\begin{equation}
A_{eq} =\left[ 
\begin{array}{cccccccc}
h_{Q_{1}} & 0 & \ldots  & h_{Q_{K}} & 0 \\ 
0 & h_{Q_{1}} & \ldots & 0 & h_{Q_{K}}%
\end{array}%
\right] \quad \text{and} \quad
b_{eq} =\left[ 
\begin{array}{c}
1 \\ 
0%
\end{array}%
\right].
\end{equation}%
The constants $\left( h_{Q_{k}}\right) _{k=1}^{K}$ may be used to assign different weights to each subdivision. In the previous example, $h_{Q_{k}}$ may be set to equal the number of delivery hours in that quarter, divided by the total number of hours per year.

The above specified model ensures that the shaping coefficients are immediately without arbitrage, since the multidimensional constrained fit removes arbitrage during the fitting process itself. Hence, the arbitrage problem is solved quite elegantly by extending the dimension of the fitting problem.

In what follows, a novel and robust way to estimate the corresponding parameters will be introduced. In order to construct an estimator suitable for the practical purposes described in the previous sections, it should satisfy the following
three requirements. At first, it should be tractable and fast in computation. The shaping
models are often used in real time to manage trading books containing
derivatives positions. Long calibration procedures can cause delays that
miss the actual market movements altogether. 

Secondly, it is important that
all parameters have an intuitive interpretation as one would typically want
to use parameter shifts in order to stress test the portfolio. This requirement mathematically boils down to the fact that the estimates have to be obtained as a set of estimates from a single estimation procedure. This can most readily be achieved by describing the regression model as a multivariate constrained model, in which each of the individual dependent variables reflect the time span of lower granularity (e.g., if the target is to estimate quarterly coefficients from annual futures, then the multivariate model has a four dimensional dependent variable in which each quarter is a column). Finally, it
is equally important that the results are robust and stable. If the fitting
procedure is sensitive to outliers, it will induce
fluctuations in the valuation of the trading books that are unmanageable.
Stable results also mean that for fixed parameters, but with changing
market prices, the highly granular prices only change in relation to the
actual market movements and not due to instabilities of the fitting model as
is the case with some smoothing algorithms. 

The above trio of requirements narrows down the theoretical set of options one has for constructing mathematical estimators. 
First of all, it is proposed to estimate the parameters using a multivariate regression model under the constraints that the coefficients are linked by some fundamental (linear) relationships. These built-in constraints allow to fit all model parameters simultaneously, whereas traditionally they would be treated separately. Secondly, a robust regression model is used to obtain more stable results. The goal of robust statistics is to develop data analytical methods which are resistant to outlying observations in the data. 
A regression data set can have different types of outliers. An observation that does not follow the linear pattern of the majority of the data but whose predictor variables are not outlying, is called a vertical outlier, whereas a point whose predictor is outlying is called a leverage point (such a point is a bad leverage point when it does not follow the pattern of the majority; otherwise, it is not harming the fit and hence a good leverage point). 
Applying a robust method is crucial for modeling derivatives pricing, since too much model uncertainty may lead to unstable results and large unhedgeable fluctuations in the trading books.

Summarizing, the estimator needs to be a constrained generalization of a robust multivariate regression estimator with high breakdown point, but yet computationally efficient. Keeping these considerations in mind, a straightforward choice is to construct a constrained generalization of robust M-regression \citep{MMreg} with highly robust starting values. The proposed multivariate constrained robust M-regression (MCRM) method is a constrained version of robust M-regression. Being a constrained robust M-estimator, the new method inherits several beneficial properties from the class of robust M-estimators (for an introduction to robust statistics, M-estimators and M-regression, we refer to \citet{Rousseeuw86-03} and \citet{Maronna}). At first, it is robust against both vertical outliers and leverage points, which guarantees stable behavior in case of isolated sparks or dips in the market. In contrast to alternatives such as MCD regression \citep{MCDreg} or LTS regression \citep{LTSreg}, M-estimators have the advantage that they are equivalent to (a subset of) iteratively reweighted least squares regression (IRLS), see \citet{IRLS,HollandWelsch77}. Therefore they can be computed very efficiently in an iterative reweighting scheme, given that highly robust estimators for the starting values are chosen. Moreover, M-estimators do algorithmically not imply any form of subset selection, as opposed to both aforementioned alternatives. This makes the proposed method computationally efficient and suitable for big data evaluation and intensive backtesting. 

The above arguments are a solid basis to assume that the proposed model is more adequate for shaping the curve than the traditional approach. In the next section, the equivalence between the M-regression and its iteratively reweighted counterpart is highlighted and a practical algorithm for the proposed methodology is presented. Its performance on real data from the German power market is shown in Section \ref{Sec:example}. 

\section{MCRM: technical details and an efficient algorithm\label{Sec:algo}}
Considering the motivation described in the previous Section, the objective to be achieved can mathematically be formulated as 
a cost function. That objective function consists of two parts: the regular
least squares term and the constraint term. Note that there are no
restrictions on the least squares term and that it does not need modification so as to be a robust
approach. Given that the sample consists of $N$ cases, such that $i \in \{1,\ldots,N\}$, the cost function is given by:
\begin{eqnarray}
Cost &=&Cost_{ls}+Cost_{c} \notag \\
&=&\sum_{k=1}^{K}\sum_{i=1}^{N}w\left( \frac{\hat{r}_i}{\hat{\sigma}}\right)   \notag \\
& &+\alpha \left( N\right) \sum_{j=1}^{M}\left(
\sum_{k=1}^{K}A_{eq}(j,k)\gamma _{k}-b_{eq}(j)\right) ^{2}, \label{eq:CRIT}
\end{eqnarray}%
where $\hat{r}_i = \hat{Y}_{i}-\hat{A}_{k}X_{i}-B_{k}$ are the regression residuals and $\hat{\sigma}$ denotes an estimate of residual scale. The function  $w(\cdot )$ can be any reasonable weight function and $\alpha \left(
N\right) $ is a scaling weight, which should scale with the number of
observations such that the constraint term does not disappear if the data set
becomes big. It is also possible to exclude the constraint from the error
term and to perform a constrained minimization. However, this makes robust fitting more tedious and is therefore not pursued \citep{Rousseeuw86-03,Huber11}. The weight function needs to be positive, zero if the error is zero,
symmetric and non-decreasing.

In the case of ordinary least squares (OLS) regression this weight function is given by $%
w\left( x\right) =x^{2}.$ In order to make the regression robust to vertical outliers, it suffices to use a bounded loss function. For instance, one could use the bi-square weight function, whose explicit
form is given by:%
\begin{equation}
w(x)=\left\{ 
\begin{array}{lcr}
{k^{2}}/{6}\cdot (1-(1-{x^{2}}/{k^{2}})^{3}) & for & |x|\leq k \\ 
{k^{2}}/{6} & for & |x|>k%
\end{array}%
\right. \mathbf{,}  \label{Ref:BiSquare}
\end{equation}

In order to minimize the cost function above, one needs to solve the set of
equations: 
\begin{equation}
\begin{cases}\label{eq:mincost}
\dfrac{\partial Cost}{\partial A_{k}}=0, \\ 
\dfrac{\partial Cost}{\partial B_{k}}=0%
\end{cases}
\end{equation}

Denote the derivative of the weight function as $\psi (x)=w^{\prime
}(x)$ and furthermore, rescale this into $\omega (x)=\psi (x)/x$. Then the
derivatives of the left-hand part (least squares part) of the cost function
are expressed as: 
\begin{equation}
\left\{ 
\begin{array}{l}
\dfrac{\partial Cost_{ls}}{\partial A_{k}}=-\sum\limits_{i=1}^{N}\omega
(Y_{i}-A_{k}X_{i}-B_{k})\cdot X_{i}\cdot (Y_{i}-A_{k}X_{i}-B_{k}) \\ 
\dfrac{\partial Cost_{ls}}{\partial B_{k}}=-\sum\limits_{i=1}^{N}\omega
(Y_{i}-A_{k}X_{i}-B_{k})\cdot (Y_{i}-A_{k}X_{i}-B_{k})%
\end{array}%
\right.  \label{Ref:LS}
\end{equation}

Equation \eqref{Ref:LS} illustrates that the multivariate regression is equivalent to a set of independent estimates in each dimension of the dependent variables, provided that there are no
constraints in the cost function. 

Optimizing Equations \eqref{eq:CRIT} for the constraint terms , yields:
\begin{equation}
\left\{ 
\begin{array}{l}
\dfrac{\partial Cost_{c}}{\partial A_{k}}=2\alpha \left( N\right)
\sum\limits_{j=1}^{M}A_{eq}(j,2k-1)\cdot \left(
\sum\limits_{l=1}^{K}A_{eq}(j,l)\gamma _{l}-b_{eq}(j)\right) \\ 
\dfrac{\partial Cost_{c}}{\partial B_{k}}=2\alpha \left( N\right)
\sum\limits_{j=1}^{M}A_{eq}(j,2k)\cdot \left(
\sum\limits_{l=1}^{K}A_{eq}(j,l)\gamma _{l}-b_{eq}(j)\right)%
\end{array}%
\right.
\end{equation}

If the weight function is given by the linear regression weight function, $%
\left( \ref{Ref:LS}\right) $ is a set of linear equations, making the
solution explicit. For the bi-square weight function $\left( \ref%
{Ref:BiSquare}\right) ,$ the problem becomes non-linear but as mentioned above, this problem can be solved by an iterative reweighting scheme (see e.g. \citet{HollandWelsch77,StreetCarrollRuppert88}).

Note that by including the constraints into the objective function, the
algorithm attempts to include them, but does not necessarily succeed.
However, by setting the weight $\alpha \left( N\right) $ appropriately, one
can put more weight on this part of the fit. It is clear that $\alpha \left(
N\right) $ needs to be an increasing function of $N$ in order to avoid the
constraint term to get relatively weakened if one adds more data points to
the problem. Furthermore, $\alpha \left( N\right) $ needs to be scaled
appropriately with the value of the data, such that a different unit of
measure would not change the fit. A straightforward and highly robust function satisfying both prerequisites is\footnote{As mentioned above, by putting the constraint terms into the objective
function, the restriction becomes a soft restriction and some care needs to
be put into ensuring that the resulting parameter estimates indeed don't
violate the non-arbitrage constraints. However, it turns out that if the
input data already contain the criterion, performing the fit this way,
works well. 
If $\alpha(N)$ would be chosen too small, the resulting fit would show arbitrage and one may simply recalculate the fit with a higher value of $\alpha(N)$.}:%
\begin{equation}
\label{eq:alpha(N)}
\alpha \left( N\right) =N\cdot Q_n(\hat{Y}),
\end{equation}
where the robust estimator of scale $Q_n$ was defined in \citep{ChristopheQn}. 
Denote the individual case weights by $\omega
_{i}= w (r_{i}/\sigma)=w \left( \frac{Y_{i}-\gamma \bar{X}%
_{i}}{\sigma}\right) $, where $\bar{X}=(X,1)$ denotes a matrix consisting of the independent variable and a column of ones to estimate an intercept term. For
the bi-square weight function, the rescaled derivative function is given by: 
\begin{equation}
\omega_{\mathrm{Biweight}} (x)=\left\{ 
\begin{array}{lcr}
(1-\left( {x}/{k}\right) ^{2})^{2} & for & |x|\leq k \\ 
0 & for & |x|>k%
\end{array}%
\right. .
\end{equation}

Various alternatives to the derivatives of the bisquare function for iterative downweighting function exist. For these purposes, Hampel's redescending weighting function \citep{Hampel} possibly yields the best tradeoff between mathematical elegance and interpretability: it is a continuous function that gradually attributes lower weights to cases that are further away from the bulk of the data, and the individual cutoff points can be chosen to correspond to quantiles of an assumed probability distribution, such as the standard normal. The Hampel function is given by:
\begin{equation}
\label{eq:hampelf}
\omega_{\mathrm{H}}(x) = \left\{\begin{matrix}
1 & & |x| \leq a \\
\frac{a}{|x|} & & a < |x| \leq b \\
\frac{r - |x|}{r -b} \frac{a}{|x|} & \mbox{if} & b < |x| \leq r \\
0 & & r < |x| . \end{matrix} 
 \right.
\end{equation}

A sensible choice for the parameters $a, b$ and $r$ are the $0.95$, $0.975$ and $0.99$ quantiles of the normal distribution. For all calculations shown in the example section, downweighting was performed by the Hampel function. 
In order to obtain a fully robust iterative reweighting procedure, two issues still need to be addressed: (i) it needs robust starting values and (ii) it relies on a scale estimate to be plugged into \eqref{eq:CRIT}. Of course, both the estimates of starting values and internal scale should be highly robust and preferably fast in terms of computation. In order to fulfil the above criteria, a good scale estimator to plug in is the $Q_n$ estimator, but of course alternatives exist.
In order to obtain robust starting values, a coarse, but highly robust measure of outlyingness can be used to determine an initial set of weights $\omega_i^{(1)}$. These initial weights are computed both in the $X$ and $Y$ dimensions, yielding an initial vector of case weights $\omega^{(1)} = \sqrt{\omega_X \omega_Y^{(1)}}$. The data are then transformed to weighted data $X_{\omega} = \Omega^{(1)}X$ and $Y_{\omega} = \Omega^{(1)}Y$, with $\Omega^{(1)}$ a diagonal matrix with diagonal elements $\omega_{i}^{(1)}\in[0,1]$ for $i\in\{1,...,N\}$.  A first regression estimate is obtained from these weighted data, yielding a first iteration estimate residuals $\hat{r}_i = Y_{\omega,i}-\hat{A}^{(1)}X_{\omega,i}-\hat{B}^{(1)}$. Based on these residuals, now an updated set of case weights can be computed: $\omega^{(2)} = \sqrt{\omega_X \omega_Y^{(2)}}$. This procedure is repeated until convergence of the $\hat{B}^{(k)}$. In more detail, the algorithm goes as follows\footnote{A Matlab implementation of the algorithm may be downloaded on the publications section of the webpage of our research group: \url{https://wis.kuleuven.be/stat/robust}. }: 

\fbox{\begin{minipage}{1\textwidth}
{\bf Algorithm: Multivariate constrained M-regression}
\newline $X$ and $Y$ denote robustly centered data (by column-wise median) consisting of $N$ cases.
\begin{enumerate}
\item Calculate initial case weights $\omega_i^{(1)}$ (for $i=1,\ldots, N$):
\begin{itemize}
\item Calculate distances for $x_i$ ($i$th row of $X$) and $y_i$ ($i$th row of $Y$):
\begin{align*}
d_i^Y&=\frac{\| y_i \|}{\mathrm{med}_h \| y_h \|} \quad \text{ and }\\
d_i^X&=\frac{| x_i |}{c\hspace{0.07cm}\mathrm{med}_h | x_h |} \quad \text{ for } h \in \{1,...,N\}
\end{align*}
where $c=1.4826$ for consistency of the MAD at the normal distribution.
\item Transform distances to block weights through the Hampel function, e.g. $\omega_X(x)=\omega_Y(x)=\omega_{\mathrm{H}}(x)$.
\item Define initial weights $\omega_i^{(1)}=\sqrt{\omega_X(d_i^X)\omega_Y(d_i^Y)}$.
\end{itemize}
\item Iteratively reweight for $k=1$, until convergence of the $\hat{B}^{(k)}$: 
\begin{itemize}
\item Construct diagonal matrix $\Omega^{(k)}$ with diagonal elements $\omega_{i}^{(k)}\in[0,1]$ for $i\in\{1,...,N\}$.
\item Weight data:
\begin{equation*}
\begin{array}{ll}
X_\omega&=\Omega^{(k)} X\\
Y_\omega&=\Omega^{(k)} Y
\end{array}
\end{equation*}
\item Minimize the cost function as in \eqref{eq:mincost} based on the weighted data $X_{\omega}$ and $Y_{\omega}$, yielding coefficients $\hat{A}^{(k)}$ and $\hat{B}^{(k)}$. Based on these coefficients, compute predicted response $\hat{Y}_{\omega}$.
\item Calculate weights for the updated responses $\hat{Y}_{\omega}$.
\begin{itemize}
\item Calculate distances for the robustly centered and scaled residuals 
for $i \in \{1,...,N\}$:
\begin{equation*}
d^{r}_{i} = \frac{| Y_{\omega,i} - \hat{Y}_{\omega,i} - \mathrm{med}_h(y_{\omega,h} - \hat{y}_{\omega,h}) |}{c\hspace{0.07cm} \mathrm{med}_{h} | Y_{\omega,h} - \hat{Y}_{\omega,h} - \mathrm{med}_\ell(y_{\omega,\ell} - \hat{y}_{\omega,\ell}) |}
\end{equation*}
\item Update weights $\omega_{i}^{(k)}=\sqrt{\omega_X(d_i^X) \omega_Y(d^{r}_{i})}$.
\end{itemize} 
\end{itemize}

\item Denote estimates of the final iteration by $\hat{A}$ and $\hat{B}$.
\end{enumerate}
\end{minipage}}

\section{Application of MCRM to German power market\label{Sec:example}}

As in Section \ref{Sec:traditional}, for every year $j$ from 2004 until 2015 data points of the form
\begin{equation*}
(Cal_{j}(t_{i});Q_1-j(t_{i}),Q_2-j(t_{i}),Q_3-j(t_{i}),Q_4-j(t_{i}))
\end{equation*}%
for individual quotation dates $t_{i}$ are obtained from the German power market. It has already
been indicated in Section \ref{Sec:traditional} that selecting a particular dataset as a representative set actually comes down to
taking a view on the market. In a competitive environment, one would most
likely try to translate fundamental forecasts about supply and demand curves
for electricity into the constraints for the coefficients.

A technical difficulty is that it is hard to get reliable data for all
contracts simultaneously. For example, the fourth quarter will start trading
liquidly much after the first quarter of a given year. And of course, at
some point the first quarter has expired and no longer quotes while the
remaining quarters are still alive. The main challenge is to have a method
that is good enough to cope with this problem and the traditional method,
described in Section \ref{Sec:traditional} is vulnerable to this.

Rather than performing individual regression models, the MCRM method ties up the
regression models through the constraints. This does not only remove the
arbitrage in the fit, but also strengthens the data naturally. By applying a
robust estimator, this effect gets improved even further. If there is a weak
quote in any of the quarters, the non-arbitrage relationship ensures that
the fit does not get affected too much by it.

\begin{figure}[!h]
\begin{center}
\includegraphics[width=9cm]{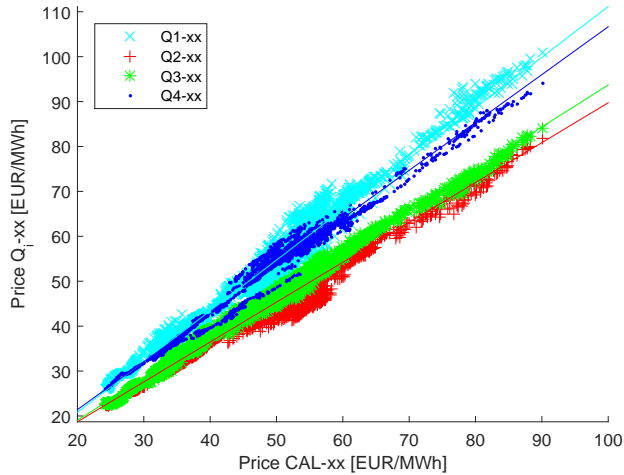}
\caption{\label{Fig:datareg} The quotation prices for each of the four quarters versus the calender year price. The data contains $Q_i-j(t_k)$ for all quotation dates $k=1,\ldots, N_{i,j}$, each quarter $i=1,\ldots, 4$ and all forward years $2004,\ldots, 2015$. A robust regression line has been added for each quarter.}
\end{center}
\end{figure}

In Figure \ref{Fig:datareg}, the quotation prices for each of the four quarters versus the calendar year price are shown for all forward years from 2004 until 2015 from the German power market. For each of the quarters, a regression line is estimated using a robust regression analysis. It is clear that not all regression lines (also added in Figure \ref{Fig:datareg}) pass through the origin and hence an intercept term might be wrongly omitted from model $\left( \ref{ShapingModel}\right)$. 
This was confirmed using Wald-type inference (see \citet{CrouxDhaeneHoorelbeke08,KollerStahel11} for more information) as the intercept term was found to be statistically significant for each quarter\footnote{The intercept terms are also significantly different from zero when using classical least squares estimation and hypothesis testing.}. Intercept therms were therefore always included in the subsequent analysis.

The shaping coefficients obtained using our robust affine MCRM methodology are shown in Table \ref{coefs:years}. For comparison, shaping coefficients were added for non-robust multivariate constrained regression. These results are obtained by setting all case weights equal to one. It can be easily checked that both results are arbitrage-free. However, subtle differences between the fitted coefficients are noticeable. 

\begin{table}[!tb]
\centering
\begin{tabular}{@{}lllllllllll@{}}
\toprule
                     &  & \multicolumn{4}{c}{slopes}                                                                                                &  & \multicolumn{4}{c}{intercepts}                                                                                            \\ \cmidrule(lr){3-6} \cmidrule(l){8-11} 
                     &  & $A_1^{YtQ}$ & $A_2^{YtQ}$ & $A_3^{YtQ}$ & $A_4^{YtQ}$ &  & $B_1^{YtQ}$ & $B_2^{YtQ} $& $B_3^{YtQ}$ & $B_4^{YtQ}$ \\ \cmidrule(r){1-1} \cmidrule(lr){3-6} \cmidrule(l){8-11} 
classical  &  & 1.146                       & 0.857                       & 0.926                       & 1.071                       &  & -2.409                      & 1.830                       & 0.610                       & -0.030                      \\
MCRM                 &  & 1.121                       & 0.875                      & 0.921                       & 1.083                       &  & -1.604                      & 1.406                       & 0.930                       & -0.732                      \\ \bottomrule 
\end{tabular}
\caption{The shaping coefficients and intercept terms fitted using classical constrained regression and MCRM for the German power market data between 2004 and 2015.}
\label{coefs:years}
\end{table}

\begin{figure}[tb]
	\centering
		\includegraphics[width=0.50\textwidth]{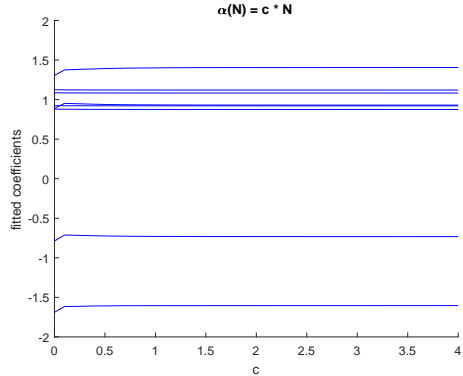}
	\caption{Plot of the fitted regression coefficients in function of the penalty $\alpha(N)$. The form of $\alpha(N)$ is chosen to be $c * N$, with $c$ varying from 0 to 4. }
	\label{fig:Sensitivity_alpha}
\end{figure}

To gauge the impact of the parameter $\alpha(N) = c\cdot N$, the obtained coefficients $A^{YtQ}_k$ and $B^{YtQ}_k$ are plotted as a function of $c$ in Figure \ref{fig:Sensitivity_alpha}. For small numbers of $c$, the term in the cost function regulating the contribution due to the constraints is not strong enough and the resulting shape coefficients are not arbitrage free. However, the estimated coefficients stabilize very rapidly as $c$ increases and the non-arbitrage condition is satisfied for $c > 2.5$. For the German power market data analyzed in this Section, $Qn(Y) = 12.49 > 2.5$. Therefore, the value of $\alpha(N)$ chosen in Equation \ref{eq:alpha(N)} leads to a solution satisfying the constraints. Furthermore, Figure \ref{fig:Sensitivity_alpha} also illustrates that the procedure is not very sensitive for the choice of $\alpha(N)$ and one may easily verify, after fitting, whether the constraints have been satisfied. If this would not be the case, one can simply increase $\alpha(N)$ to obtain a solution satisfying the constraints. 

\begin{figure}
\centering
\begin{subfigure}{.5\textwidth}
				\begin{center}
				\includegraphics[width=\textwidth]{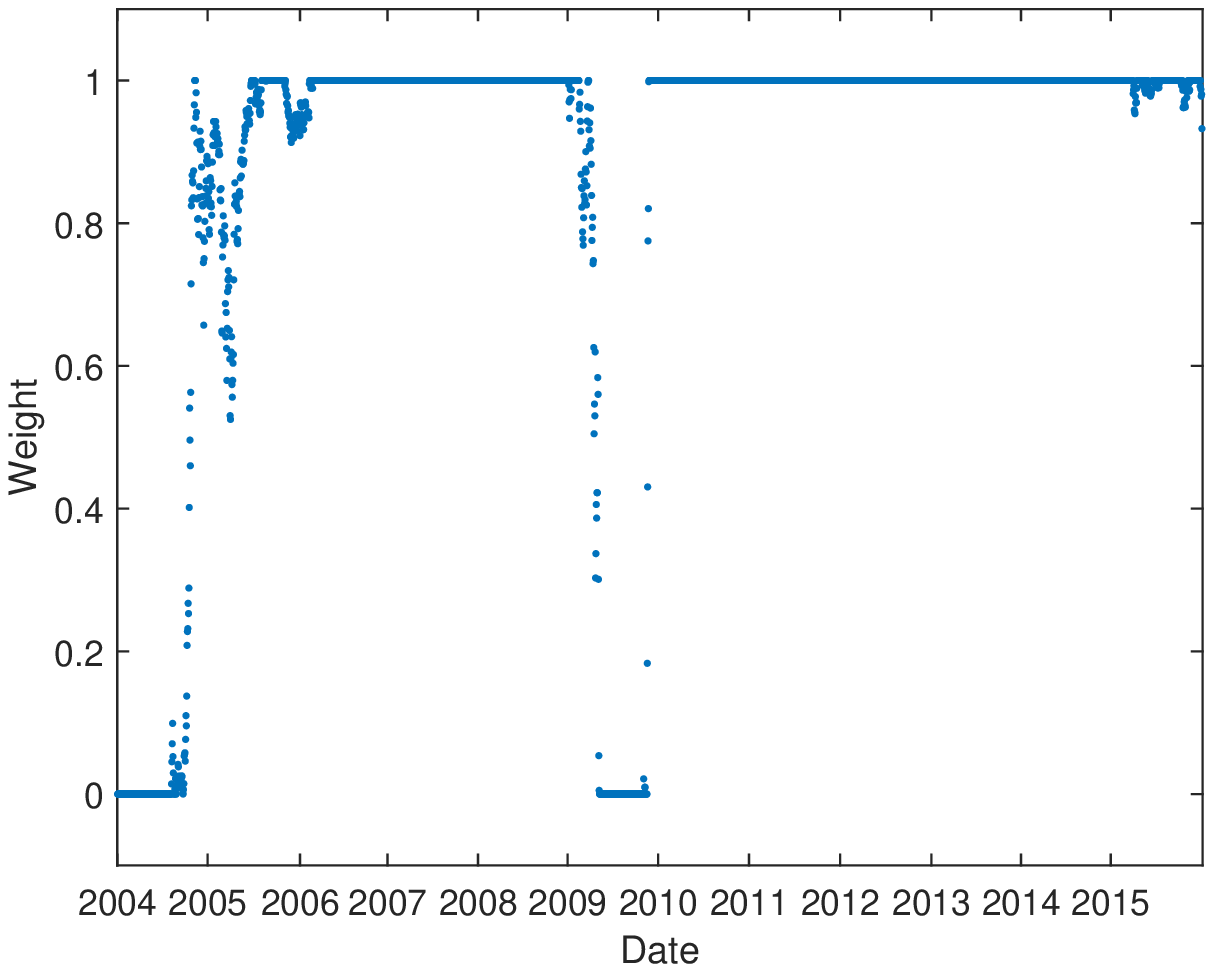}
				\caption{}
				\label{Fig:outliersYearly_weights}
				\end{center}
\end{subfigure}%
\begin{subfigure}{.5\textwidth}
				\begin{center}
				\includegraphics[width=\textwidth]{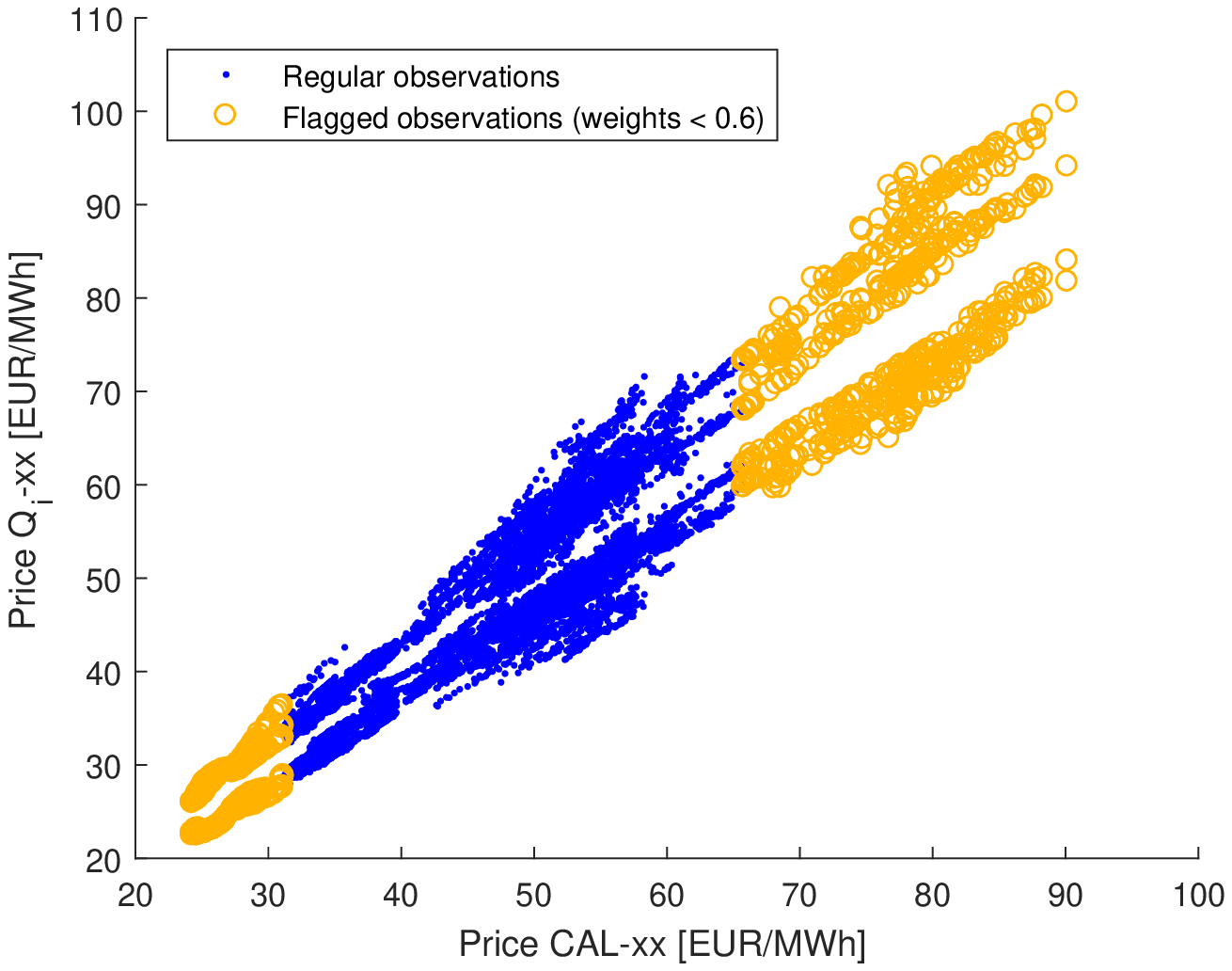}
				\caption{}
				\label{Fig:outliersYearly_prices}
				\end{center}
\end{subfigure}
				\caption{The final weights $\omega_i$ for every observation in the German power market data ($i=1,\ldots,N$).}
\end{figure}

A benefit of the robust methodology is that the final weights can be studied to correctly and automatically detect atypical observations in the data. Figure \ref{Fig:outliersYearly_weights} shows the weight for every observation date in the data. The smaller the weight, the more atypical an observation is (a weight of one is attributed to the regular observations). Notably, series of downweighted cases correspond to time slots in which the markets indeed behaved atypically, such as the 2009 crisis. 

To visualize these results, the quotation prices for each of the four quarters are again plotted versus the calendar year price, but observations with a weight smaller than 0.6 are now marked by orange circles, whereas regular observations are marked by blue dots (Figure \ref{Fig:outliersYearly_prices}). From this it is clear that the flagged outliers are indeed deviating from the pattern of the majority of the data.

To gauge the impact of these outliers on the fit, consider the following measures:
\[\text{MeanAE} = \frac{1}{Kn} \sum_{i=1}^{n} \sum_{j=1}^{K}|Q_j^i - \hat{Q}_j^i|, \quad \text{MedAE} = \mbox{med}_{i} \frac{1}{K} \sum_{j=1}^{K}|Q_j^i - \hat{Q}_j^i|,\]
\[\text{MeanSE} = \frac{1}{Kn} \sum_{i=1}^{n} \sum_{j=1}^{K}(Q_j^i - \hat{Q}_j^i)^2, \quad \text{MedSE} = \mbox{med}_{i} \frac{1}{K} \sum_{j=1}^{K}(Q_j^i - \hat{Q}_j^i)^2,\]

comparing the estimated quarterly price $\hat{Q}_j^i$ with the actual quarterly price $Q_j^i$. Moreover,the results are compared to the results obtained by Fleten and Lemming's method. The results are shown summarized in Table \ref{tab:dataResul}.

\begin{table}[!tb]
\footnotesize	
\centering
\begin{tabular}{@{}llllllllllll@{}}
\toprule
                    & &  & \multicolumn{4}{c}{in-sample} &  & \multicolumn{4}{c}{out-of-sample}\\ 
										\cmidrule(lr){4-7} \cmidrule(l){9-12} 
										& &  & \multicolumn{2}{c}{AE} & \multicolumn{2}{c}{SE} &  & \multicolumn{2}{c}{AE} & \multicolumn{2}{c}{SE}\\ 
										\cmidrule(lr){4-5} \cmidrule(lr){6-7} \cmidrule(l){9-10}  \cmidrule(l){11-12}
                    & &  & mean & med & mean & med &  & mean & med & mean & med  \\
\midrule
\multirow{3}{*}{\rotatebox[origin=c]{90}{quarterly}}&F\&L     & &	 1.282 & 0.891 & 3.638 & \textbf{0.973} & & 0.256 & 0.236 & \textbf{0.092} & 0.062\\ [0.2ex]
&classical& &   1.176 & 0.965 & \textbf{2.700} & 1.172 & & 0.296 & 0.269 & 1.149 & 0.106\\	 [0.2ex]
&robust   & &   \textbf{1.172} & \textbf{0.930} & 2.818 & 1.126 & & \textbf{0.232} & \textbf{0.198} & 0.094 & \textbf{0.054}\\	 [0.2ex]	
\midrule 
\multirow{3}{*}{\rotatebox[origin=c]{90}{hourly}}&F\&L     & &	 5.124 & 4.460 & 62.94 & 30.27 & & 4.629 & 4.283 & 36.92 & 26.16 \\  [0.2ex]
&classical& &   5.025 & 4.384 & \textbf{57.37} & 29.36 & & 5.066 & 4.867 & 41.37 & 33.07 \\	  [0.2ex]
&robust   & &   \textbf{5.025} & \textbf{4.080 }& 60.45 & \textbf{26.16} & & \textbf{4.505} & \textbf{4.099} & \textbf{35.67} & \textbf{25.02} \\	 [0.2ex]
												
										\bottomrule 
\end{tabular}
\normalsize
\caption{Summary statistics of the in-sample and out of sample performance of the different estimators for the quarterly and hourly German power market data. }
\label{tab:dataResul}
\end{table}

Within the sample the methods have been estimated from, the results for the classical and the robust fit are comparable. The method of Fleten and Lemming, however, generally performs worse. Given the context of the data problem, also the prediction performance of the different methods is important. Therefore, the first 44 quotation dates of 2016 from the German power market are used as an out-of-sample test set. The different performance measures summarized in Table \ref{tab:dataResul} are also evaluated for these out-of-sample data. When applied to out-of-sample data, the robust method clearly shows the best performance, thereby illustrating superior stability over both the classical approach and Fleten and Lemming's method.

As a next example, hourly prices of weekdays from January, 1 2009 until December, 31 2014, are investigated. The testing set consists of weekdays from January,1 until November 26, 2015. Again, robust MCRM, classical MCRM (with all weights set to 1) and the method of Fleten and Lemming are compared. The results of the in-sample and out-of-samples test are given in Table \ref{tab:dataResul}. 

\begin{figure}[htb]
    \centering
    \begin{minipage}{0.5\textwidth}
      \includegraphics[width=\textwidth]{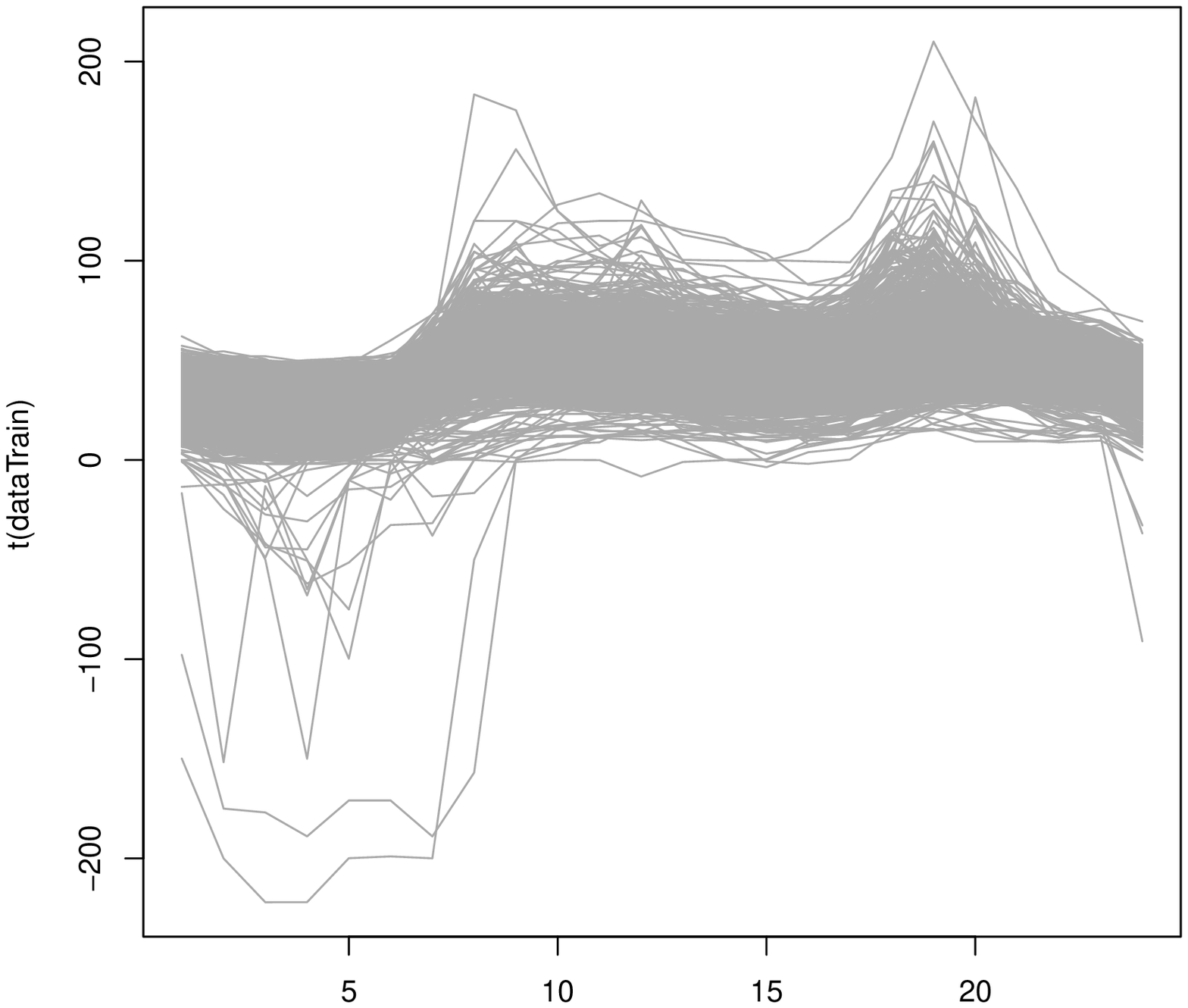}
    \end{minipage}%
		\hfill
    \begin{minipage}{0.5\textwidth}
				\includegraphics[width=\textwidth]{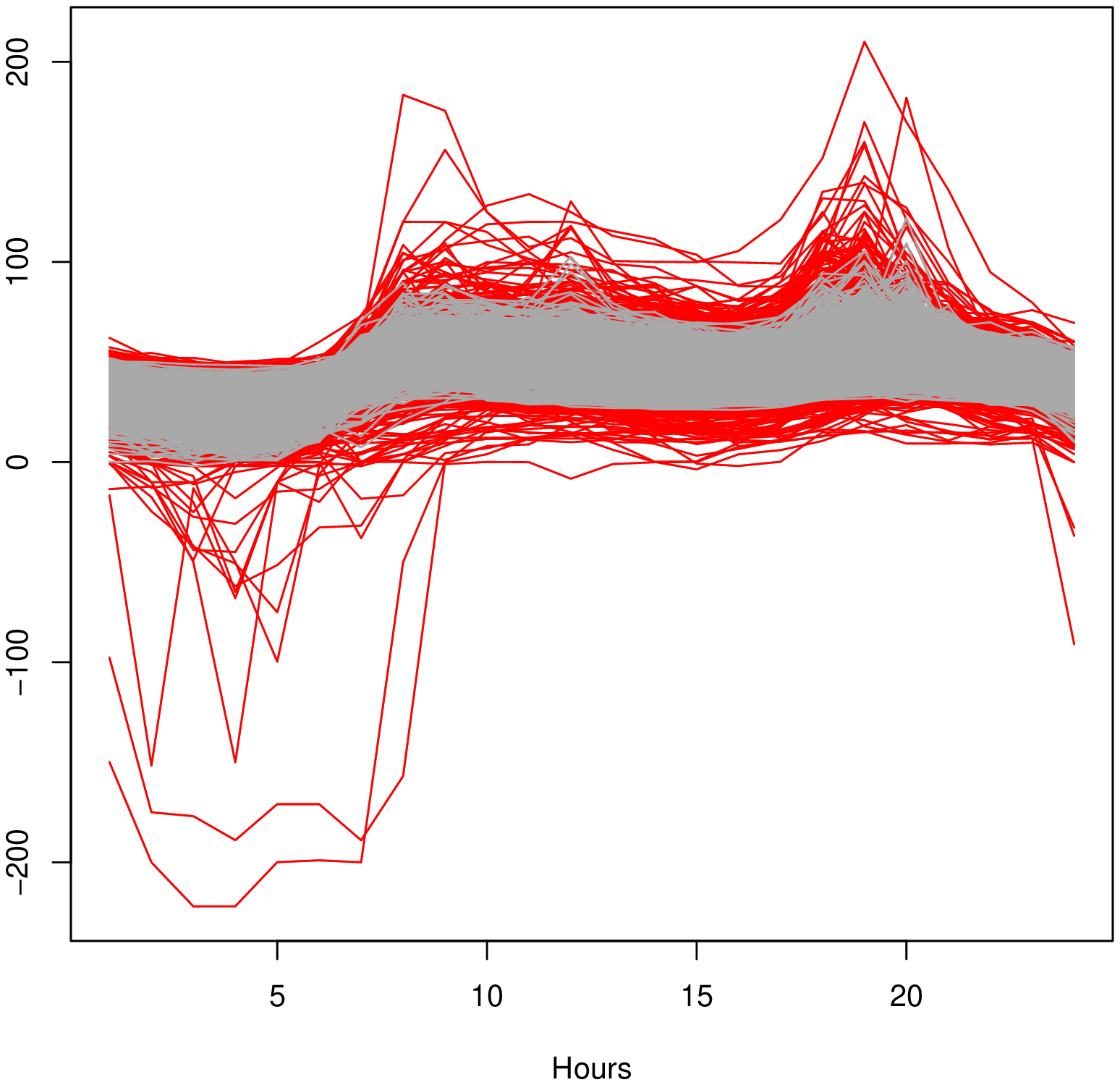}
	  \end{minipage}%
		 \caption{Settlement prices for each of the 24 hours for the weekdays from January 1, 2009 to December 31, 2014 in grey (left). On the right hand side, daily patterns recieving a weight $\omega_i < 1$ from the robust method, are marked in red. }
		\label{fig:hourlyData}
\end{figure}

For the in-sample results, one can see that the proposed multivariate constrained regression method has a small benefit over the method of Fleten and Lemming. The classical fit has a slightly lower MeanSE. This may easily be understood as the classical regression method tries to provide the best fit for all data points, whereas the robust method will, by design, fit the majority of the data well, but not the outliers. Data points flagged as outliers will, therefore, be fitted badly, inflating error measures based on averaging. 
This is clearly illustrated by by the large difference between the MeanSE and MedSE. As per the latter criterion, the MCRM estimator is optimal, confirming its robustness. Switching attention to the out-of-sample prediction errors, the classical regression method performs worse. This is a clear indication that although the in-sample fitted errors were lower, these results were influenced by deviating patterns in the data. On the contrary, the results for the proposed robust methodology are highly competitive and the errors are lower than those for both the Fleten and Lemming method and the classical constrained regression method. In Figure \ref{fig:hourlyData}, all hourly patterns flagged as suspicious by the robust method have been colored red. Indeed, it can be seen that observations deviating from the main trend have been flagged. 

The proposed robust MCRM method can clearly outperform classical estimators in scenarios were deviating observations are presented in the data. It provides an objective way for flagging influential data points and provides better prediction performance for contaminated training data. Another main benefit offered by robust regression is that it is reliable in all market scenarios considered: as long as prices are mid-range, it will yield estimates that are virtually equal to the classical ones, whereas in extreme price environments, be it at high or low price levels, the robust estimates will represent the actual shape more accurately. Note that especially in these explicitly high and low priced market scenarios, individual outlying days do occur more frequently. In low priced scenarios, the actual shape estimates from the robust method are more pronounced, which enables the trader to identify hedging opportunities that would not be detected by the classical method. Likewise, the robust method will not overestimate hedging potential in high priced markets. 

With the regulator requiring more and more detailed insights into the balance sheets, stress tests are becoming a more commonly accepted tool to analyze positions and risks. The benefit of having an affine relationship is that if one wants to
stress-test a trading book, it is quite natural to shift the constant $B_{i}$
on the highest granular level because one can exactly understand what the price
effect is. Note that the shaping coefficients are applied in sequence to
find the hourly price, e.g.%
\footnotesize\begin{eqnarray*}
&&P\left( t_{0},hour=3,day=Sat,Month=4,Year=2014\right) \\
&=&A_{3}^{DtH}\left( A_{SAT}^{MtD}\left( A_{Apr}^{QtM}\left(
A_{2}^{YtQ}\cdot F\left( t_{0},Cal-14\right) +B_{2}^{YtQ}\right)
+B_{Apr}^{QtM}\right) +B_{SAT}^{MtD}\right) +B_{3}^{DtH}
\end{eqnarray*}
\normalsize

At a certain point in time, the contract $Q_1-14$ will start to be actively
traded and one no longer has to rely on the estimate from the shaping
coefficients. The way this is dealt with in the traditional method, is to fix
the price of $Q_1$ and to rescale the other three prices. However, the
methodology presented here can allow for a quick recalibration where the
shaping coefficients for $Q1$ are fixed such that the market price is matched.
The robust regression will then estimate the coefficients for the remaining
3 quarters, while still ensuring that the non-arbitrage relationships are
honoured. This is a lot more stable and sensible than blindly reaveraging.

\section{Further applications and conclusions\label{Sec:conclusion}}
In this paper, a multivariate constrained robust M-regression method has been proposed, as well as a practical algorithm to obtain these robust estimates under a particular set of constraints. Its development has been inspired by and directly applicable to (European) electricity markets and model arbitrage is ruled out at a very elementary level. The benefits of the method have been extensively discussed and its proper functionality has been illustrated on real pricing data from the German power market. 

Based on results from historical data sets, the necessity to include an intercept in the model, has been justfied. Therefore, the approach proposed here yields both intercept and slope, in contrast to some more common ways to estimate shaping coefficients that involve arbitrage. Obviously, having a single
coefficient allows one to immediately see the effect of an increase of $1$
Eur/MWh for the $Cal$ price onto the Quarterly prices, whereas having an
affine relationship requires a little more attention. 
However, the interpretations are still very intuitive and this extension is
the natural next step.

Even if the practitioner still wants to use the simple scaling relationship of model $\left(\ref{ShapingModel}\right)$,
the MCRM still adds value, as one can set up the constraint matrix in $\left( %
\ref{Ref:Equality}\right) $ such that the shift coefficients are zero. This
will at least ensure a better way of removing the arbitrage in the
coefficients as the fit is executed simultaneously in all dimensions. In fact some traders prefer to only use the intercept and no slope. This can be included into the constraint matrix easily as well. 

Another advantage is that the proposed method can easily be extended to include more
fundamental information into the model. For example, when trying to
understand the shape of the forward curve for electricity, one can bring in
the fuel prices that generate electricity. The MCRM method can easily be
extended to be multidimensional on the regressors as well.

An interesting extension is to study robust inference for the proposed method. 
This might be obtained using Wald type inferences \citep{KollerStahel11} or using the robust bootstrap as in \citet{VanAelstWillems05}. Inference would provide standard errors of the robust point estimates and since the fit is executed in a
multidimensional setting, the corresponding confidence levels are part of the
multivariate distribution, rather than isolated one-dimensional confidence
levels. 

Although the developed methodology has only been applied to financial markets, it is very general and can also be interesting in other fields of research. 



\begin{thebibliography}{}

\bibitem[\protect\citeauthoryear{Adams and Van~Deventer}{Adams and
  Van~Deventer}{1994}]{adams1994fitting}
Adams, K.~J. and D.~R. Van~Deventer (1994).
\newblock Fitting yield curves and forward rate curves with maximum smoothness.
\newblock {\em Journal of Fixed Income\/}~{\em 4\/}(1), 52--62.

\bibitem[\protect\citeauthoryear{Benth, Koekebakker, and Ollmar}{Benth
  et~al.}{2007}]{benth2007extracting}
Benth, F.~E., S.~Koekebakker, and F.~Ollmar (2007).
\newblock Extracting and applying smooth forward curves from average-based
  commodity contracts with seasonal variation.
\newblock {\em Journal of Derivatives\/}~{\em 15\/}(1), 52.

\bibitem[\protect\citeauthoryear{Benth, Koekkebakker, and Ollmar}{Benth
  et~al.}{2007}]{Koekebakker07}
Benth, F.~E., S.~Koekkebakker, and F.~Ollmar (2007).
\newblock Extracting and applying smooth forward curves from average-based
  commodity contracts with seasonal variation.
\newblock {\em The Journal of Derivatives\/}~{\em 15\/}(1), 52--66.

\bibitem[\protect\citeauthoryear{Boogert and Dupont}{Boogert and
  Dupont}{2008}]{BoogertDupont06}
Boogert, A. and D.~Dupont (2008).
\newblock When supply meets demand: the case of hourly spot electricity prices.
\newblock {\em IEEE Transactions on Power Systems\/}~{\em 23\/}(2), 389--398.

\bibitem[\protect\citeauthoryear{Borak and Weron}{Borak and
  Weron}{2008}]{Borak08}
Borak, S. and R.~Weron (2008).
\newblock A semiparametric factor model for electricity forward curve dynamics.
\newblock {\em Journal of Energy Markets\/}~{\em 1\/}(3), 3--16.

\bibitem[\protect\citeauthoryear{Bunn}{Bunn}{2004}]{Bunn04}
Bunn, D.~W. (Ed.) (2004).
\newblock {\em Modelling Prices in Competitive Electricity Markets}.
\newblock Wiley Finance.

\bibitem[\protect\citeauthoryear{Caldana, Fusai, and Roncoroni}{Caldana
  et~al.}{2016}]{caldana2016electricity}
Caldana, R., G.~Fusai, and A.~Roncoroni (2016).
\newblock Electricity forward curves with thin granularity.

\bibitem[\protect\citeauthoryear{Clewlow and Strickland}{Clewlow and
  Strickland}{2000}]{ClewlowStrickland00}
Clewlow, L. and C.~Strickland (2000).
\newblock {\em Energy derivatives: pricing and risk management}.
\newblock Lacima Publ.

\bibitem[\protect\citeauthoryear{Croux, Dhaene, and Hoorelbeke}{Croux
  et~al.}{2004}]{CrouxDhaeneHoorelbeke08}
Croux, C., G.~Dhaene, and D.~Hoorelbeke (2004).
\newblock Robust standard errors for robust estimators.
\newblock Discussion paper, FEB Research Report, KU Leuven.

\bibitem[\protect\citeauthoryear{Davies and Gather}{Davies and
  Gather}{1993}]{davies1993identification}
Davies, L. and U.~Gather (1993).
\newblock The identification of multiple outliers.
\newblock {\em Journal of the American Statistical Association\/}~{\em
  88\/}(423), 782--792.

\bibitem[\protect\citeauthoryear{Eydeland and Wolyniec}{Eydeland and
  Wolyniec}{2003}]{EydelandWolyniec03}
Eydeland, A. and K.~Wolyniec (2003).
\newblock {\em Energy and power risk management: New developments in modeling,
  pricing, and hedging}, Volume 206.
\newblock John Wiley \& Sons.

\bibitem[\protect\citeauthoryear{Fiorenzani}{Fiorenzani}{2006}]{Fiorenzani06}
Fiorenzani, S. (2006).
\newblock {\em Quantitative methods for electricity trading and risk
  management: Advanced mathematical and statistical methods for energy
  finance}.
\newblock Springer.

\bibitem[\protect\citeauthoryear{Fleten and Lemming}{Fleten and
  Lemming}{2003}]{fleten2003constructing}
Fleten, S.-E. and J.~Lemming (2003).
\newblock Constructing forward price curves in electricity markets.
\newblock {\em Energy Economics\/}~{\em 25\/}(5), 409--424.

\bibitem[\protect\citeauthoryear{Geman}{Geman}{2005}]{Geman05}
Geman, H. (2005).
\newblock {\em Commodities and commodity derivatives: pricing and modeling
  agricultural, metals and energy}.
\newblock Wiley Finance.

\bibitem[\protect\citeauthoryear{Green}{Green}{1984}]{IRLS}
Green, P.~J. (1984).
\newblock Iteratively reweighted least squares for maximum likelihood
  estimation, and some robust and resistant alternatives.
\newblock {\em Journal of the Royal Statistical Society. Series B
  (Methodological)\/}, 149--192.

\bibitem[\protect\citeauthoryear{Hampel, Ronchetti, Rousseeuw, and
  Stahel}{Hampel et~al.}{2011}]{Hampel}
Hampel, F.~R., E.~M. Ronchetti, P.~J. Rousseeuw, and W.~A. Stahel (2011).
\newblock {\em Robust statistics: the approach based on influence functions},
  Volume 114.
\newblock John Wiley \& Sons.

\bibitem[\protect\citeauthoryear{Hildmann, Herzog, Stokic, Cornel, and
  Andersson}{Hildmann et~al.}{2011}]{hildmann2011robust}
Hildmann, M., F.~Herzog, D.~Stokic, J.~Cornel, and G.~Andersson (2011).
\newblock Robust calculation and parameter estimation of the hourly price
  forward curve.
\newblock In {\em 17th Power Systems Computation Conference (PSCC)}.

\bibitem[\protect\citeauthoryear{Hildmann, Kaffe, He, and Andersson}{Hildmann
  et~al.}{2012}]{hildmann2012combined}
Hildmann, M., E.~Kaffe, Y.~He, and G.~Andersson (2012).
\newblock Combined estimation and prediction of the hourly price forward curve.
\newblock In {\em Power and Energy Society General Meeting, 2012 IEEE}, pp.\
  1--8. IEEE.

\bibitem[\protect\citeauthoryear{Holland and Welsch}{Holland and
  Welsch}{1977}]{HollandWelsch77}
Holland, P.~W. and R.~E. Welsch (1977).
\newblock Robust regression using iteratively reweighted least-squares.
\newblock {\em Communications in Statistics-theory and Methods\/}~{\em 6\/}(9),
  813--827.

\bibitem[\protect\citeauthoryear{Huber}{Huber}{2011}]{Huber11}
Huber, P.~J. (2011).
\newblock {\em Robust statistics}.
\newblock Springer.

\bibitem[\protect\citeauthoryear{Huisman}{Huisman}{2009}]{Huisman09}
Huisman, R. (2009).
\newblock {\em Introduction to models for the energy markets}.
\newblock Risk Books.

\bibitem[\protect\citeauthoryear{James}{James}{2012}]{James12}
James, T. (2012).
\newblock {\em Energy markets: price risk management and trading}.
\newblock John Wiley \& Sons.

\bibitem[\protect\citeauthoryear{Kiesel, Paraschiv, and S{\ae}ther{\o}}{Kiesel
  et~al.}{2018}]{kiesel2018construction}
Kiesel, R., F.~Paraschiv, and A.~S{\ae}ther{\o} (2018).
\newblock On the construction of hourly price forward curves for electricity
  prices.
\newblock {\em Computational Management Science\/}, 1--25.

\bibitem[\protect\citeauthoryear{Koekebakker and Os~{\AA}dland}{Koekebakker and
  Os~{\AA}dland}{2004}]{koekebakker2004modelling}
Koekebakker, S. and R.~Os~{\AA}dland (2004).
\newblock Modelling forward freight rate dynamics—empirical evidence from
  time charter rates.
\newblock {\em Maritime Policy \& Management\/}~{\em 31\/}(4), 319--335.

\bibitem[\protect\citeauthoryear{Koller and Stahel}{Koller and
  Stahel}{2011}]{KollerStahel11}
Koller, M. and W.~A. Stahel (2011).
\newblock Sharpening wald-type inference in robust regression for small
  samples.
\newblock {\em Computational Statistics \& Data Analysis\/}~{\em 55\/}(8),
  2504--2515.

\bibitem[\protect\citeauthoryear{Maronna, Martin, and Yohai}{Maronna
  et~al.}{2006}]{Maronna}
Maronna, R., D.~Martin, and V.~Yohai (2006).
\newblock {\em Robust statistics}.
\newblock John Wiley \& Sons, Chichester. ISBN.

\bibitem[\protect\citeauthoryear{Paraschiv, Bunn, and Westgaard}{Paraschiv
  et~al.}{2016}]{paraschiv2016estimation}
Paraschiv, F., D.~Bunn, and S.~Westgaard (2016).
\newblock Estimation and application of fully parametric multifactor quantile
  regression with dynamic coefficients.
\newblock Available at SSRN 2741692.

\bibitem[\protect\citeauthoryear{Pilipovic}{Pilipovic}{2007}]{Pilipovic07}
Pilipovic, D. (2007).
\newblock {\em Energy Risk: Valuing and Managing Energy Derivatives: Valuing
  and Managing Energy Derivatives}.
\newblock McGraw Hill Professional.

\bibitem[\protect\citeauthoryear{Rousseeuw and Croux}{Rousseeuw and
  Croux}{1993}]{ChristopheQn}
Rousseeuw, P.~J. and C.~Croux (1993).
\newblock Alternatives to the median absolute deviation.
\newblock {\em Journal of the American Statistical association\/}~{\em
  88\/}(424), 1273--1283.

\bibitem[\protect\citeauthoryear{Rousseeuw and Leroy}{Rousseeuw and
  Leroy}{2005}]{Rousseeuw86-03}
Rousseeuw, P.~J. and A.~M. Leroy (2005).
\newblock {\em Robust regression and outlier detection}, Volume 589.
\newblock John Wiley \& Sons.

\bibitem[\protect\citeauthoryear{Rousseeuw, Van~Aelst, Van~Driessen, and
  Gull{\'o}}{Rousseeuw et~al.}{2012}]{MCDreg}
Rousseeuw, P.~J., S.~Van~Aelst, K.~Van~Driessen, and J.~A. Gull{\'o} (2012).
\newblock Robust multivariate regression.
\newblock {\em Technometrics\/}.

\bibitem[\protect\citeauthoryear{Rousseeuw and Van~Driessen}{Rousseeuw and
  Van~Driessen}{2006}]{LTSreg}
Rousseeuw, P.~J. and K.~Van~Driessen (2006).
\newblock Computing lts regression for large data sets.
\newblock {\em Data mining and knowledge discovery\/}~{\em 12\/}(1), 29--45.

\bibitem[\protect\citeauthoryear{Schofield}{Schofield}{2011}]{Schofield07}
Schofield, N.~C. (2011).
\newblock {\em Commodity derivatives: markets and applications}, Volume 570.
\newblock John Wiley \& Sons.

\bibitem[\protect\citeauthoryear{Skantze and Ilic}{Skantze and
  Ilic}{2012}]{SkantzeIlic12}
Skantze, P.~L. and M.~Ilic (2012).
\newblock {\em Valuation, hedging and speculation in competitive electricity
  markets: a fundamental approach}.
\newblock Springer Science \& Business Media.

\bibitem[\protect\citeauthoryear{Street, Carroll, and Ruppert}{Street
  et~al.}{1988}]{StreetCarrollRuppert88}
Street, J.~O., R.~J. Carroll, and D.~Ruppert (1988).
\newblock A note on computing robust regression estimates via iteratively
  reweighted least squares.
\newblock {\em The American Statistician\/}~{\em 42\/}(2), 152--154.

\bibitem[\protect\citeauthoryear{Weron}{Weron}{2007}]{Weron07}
Weron, R. (2007).
\newblock {\em Modeling and forecasting electricity loads and prices: A
  statistical approach}, Volume 403.
\newblock John Wiley \& Sons.

\bibitem[\protect\citeauthoryear{Willems and Van~Aelst}{Willems and
  Van~Aelst}{2005}]{VanAelstWillems05}
Willems, G. and S.~Van~Aelst (2005).
\newblock Fast and robust bootstrap for lts.
\newblock {\em Computational statistics \& data analysis\/}~{\em 48\/}(4),
  703--715.

\bibitem[\protect\citeauthoryear{Yohai}{Yohai}{1987}]{MMreg}
Yohai, V.~J. (1987).
\newblock High breakdown-point and high efficiency robust estimates for
  regression.
\newblock {\em The Annals of Statistics\/}, 642--656.

\end{thebibliography}
\end{document}